\documentclass[aps,prl,reprintgroupedaddress,longbibliography]{revtex4-1}
\usepackage{amsmath}
\usepackage{graphicx}
\usepackage{caption}
\usepackage{subcaption}
\usepackage{hyperref}
\usepackage{lineno}
\usepackage{tikz}
\usetikzlibrary{positioning}

\begin{document}

\title{Thermodynamic and Statistical Signatures of Modality Changes in Concentration Distributions Driven by Stochastic Switching Between Two Activity States}

\author{Aindrila Deb and Pintu Patra$^{1}$} 
\email[]{pintupatra@phy.iitkgp.ac.in}
\affiliation{$^{1}$Department of Physics, Indian Institute of Technology Kharagpur, West Bengal}
\date{\today}

\begin{abstract}
Stochastic switching between gene expression states, coupled with production and degradation dynamics, governs the accumulation of mRNA and proteins in cells. The concentrations of these accumulated entities dictate the phenotypic distribution of genetically identical cells. The underlying accumulation dynamics are well-captured by a two-state promoter switching model, with statistical and thermodynamic properties quantified via the Fano factor and entropy production rates. However, how these measures correlate with concentration distributions and their shifts under varying kinetic parameters remains largely unexplored. To this end, we use chemical master equations to study a generalized model of mRNA accumulation dynamics in the presence of stochastic switching between two activity states and state-dependent production and degradation rates. We derive exact expressions for the steady-state probability distribution and analytically compute the mean concentration, Fano factor, and entropy production rate (EPR). Simplifying these expressions, we identify contributions arising from stochastic switching rates and relaxation dynamics toward equilibrium in each activity state. Next, using our theoretical results, we characterize the variation in the Fano factor and EPR as a function of mean expression during modality changes of the distributions mediated by the variation of switching rates. We also identify the conditions in kinetic parameters that achieve the highest Fano factor and entropy production rates. Our findings establish a generalized framework for examining stochastic accumulation dynamics, clarifying how kinetic parameters dictate molecular distributions, noise, and dissipation. These insights extend readily to broader contexts coupling stochastic switching with accumulation, including protein burst dynamics, phenotype-switching-mediated drug intake, and queuing theory.
\end{abstract}

\maketitle
 
\section{Introduction}
Phenotypic heterogeneity is ubiquitous in living systems, enabling genetically identical cells to adopt distinct physiological states \cite{Elowitz2002, Raj2008, Altschuler2010}. It serves as a general mechanism for diversifying population responses to changing environments \cite{kaern2005stochasticity, Balaban2004, Kussell2005, Patra2013}. At the molecular level, phenotypic heterogeneity arises due to stochastic switching between distinct gene expression states, which dictates downstream molecular reactions, such as mRNA synthesis and protein production \cite{Paulsson2004, Karmakar2020} \cite{kaern2005stochasticity, eldar2010functional, munsky2012using}. Such stochastic switching between distinct states is widely observed across biological processes \cite{Elowitz2002, Swain2002}, including transcriptional bursting \cite{Raj2006, Golding2005}, bacterial persistence \cite{Balaban2004, Kussell2005, Patra2013}, cell-fate determination \cite{Losick2008}, epithelial-mesenchymal transitions \cite{Gupta2011, Lu2013}, and stress adaptation \cite{Acar2008, garcia2016phenotypic}. The two-state approach to gene expression dynamics provides a foundational framework for studying stochastic switching \cite{Elowitz2002, Swain2002, Raser2005, Raj2006, Hung2014}, commonly represented by the canonical telegraph model \cite{Peccoud1995}. In this model, a promoter stochastically switches between inactive (OFF) and active (ON) states, with transcription occurring only in the ON state. This model recovers the statistical features of mRNA concentration distributions, which can range from unimodal to bimodal \cite{Raj2006, IyerBiswas2009, bothma2014dynamic, cao2020stochastic, shahrezaei2008analytical, visco2008exact, Hung2014}. Further, to account for basal or leaky expression in the inactive state, the model has been generalized to allow state-dependent production rates in both states \cite{Ham2020, kepler2001stochasticity, Holehouse2026, zhang2022rate}. However, the relationship between these statistical measures and the underlying concentration distributions, an experimentally accessible quantity \cite{Raj2006, IyerBiswas2009, bothma2014dynamic, cao2020stochastic, Hung2014, hu2018eicirna}, has received comparatively less attention. Further, how statistical fluctuations and thermodynamic cost vary as the system transitions between different distributions, e.g, in tunable synthetic gene networks \cite{Hung2014}, during disease progression and evolution in a changing environment \cite{sun2020stochastic}, remains largely unexplored. 

In this work, we generalize the two-state model of gene expression to incorporate both state-dependent production and degradation rates. In conventional formulations, mRNA degradation is typically assumed to occur at a constant rate, independent of the promoter state \cite{Peccoud1995, Ham2020, kepler2001stochasticity, Holehouse2026}. However, in bacterial systems, transcription, translation, and mRNA degradation can be kinetically coupled, such that differences in regulatory and translational states may lead to differences in the effective stability of the corresponding mRNA \cite{shine2024co, trinquier2023effect, kim2026spatial}. Such coupling can arise through mechanisms including translation-dependent mRNA stability, small-RNA (sRNA)-mediated regulation, ribosome-mediated protection, and riboswitch-mediated control of mRNA decay \cite{trinquier2023effect, kim2026spatial, richards2021riboswitch, vargas2020regulation}. Motivated by these observations, we investigate how even small state-dependent differences in mRNA stability modify the statistical and thermodynamic properties of mRNA accumulation, beyond the effects arising from state-dependent transcription and promoter switching. At the same time, the constant-degradation model provides a natural null model against which state-dependent degradation can be tested, allowing one to assess changes in mRNA stability arising from secondary regulatory mechanisms \cite{park2018chemical, dattani2017stochastic}.
The generalized accumulation model is described by coupled chemical master equations, which we solve analytically to obtain the steady-state probability distribution, statistical moments, and other quantitative metrics. Specifically, we derive closed-form expressions for statistical fluctuations via the Fano factor and thermodynamic dissipation via the entropy production rate (EPR). Using these analytical results, we systematically map the changes in distribution and the corresponding Fano factor and EPR across the parameter space governing stochastic switching dynamics. Our model and results can be applied to various two-state stochastic switching processes coupled with downstream production and degradation dynamics — such as bursting protein dynamics \cite{cai2006stochastic, friedman2006linking}, intracellular drug accumulation \cite{Pu2016,roy2021persister} and transitions in cancer phenotypes \cite{kumar2019stochastic, Gupta2011, bu2013microrna, de2020shape}.

\begin{figure}[t]
    \centering
    \includegraphics[width=0.75\linewidth]{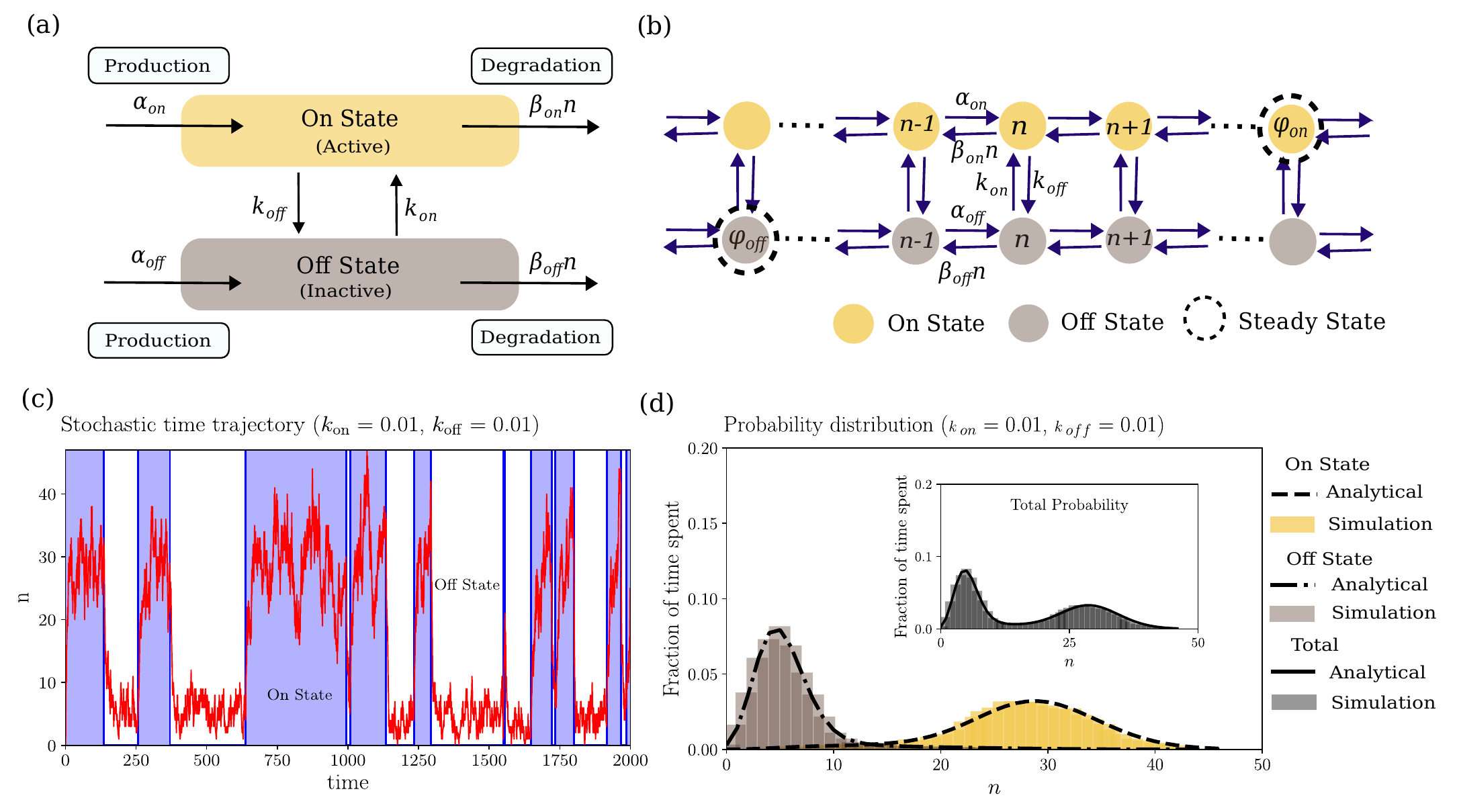}
    \caption{\textbf{Stochastic switching between two activity states coupled to accumulation dynamics}. 
\textbf{(a)} We consider a genetic state that stochastically transitions between the ON (active) and OFF (inactive) activity states and vice versa with rates $k_{\rm on}$ and $k_{\rm off}$, respectively. In each state, the molecular copy number, $n$, undergoes production at rates $\alpha_{\rm on}$ or $\alpha_{\rm off}$, and linear degradation at rates $\beta_{\rm on}n$ or $\beta_{\rm off}n$. 
\textbf{(b)} The joint state space $(S, n)$ illustrates different transitions. Horizontal transitions represent intra-state production and degradation processes, while vertical transitions represent discrete switching between the two activity states. The dashed circles show the local steady-state accumulation levels $\phi_{\rm on}$ and $\phi_{\rm off}$ for the ON and OFF states, respectively (in the absence of stochastic switching). 
\textbf{(c)} A representative stochastic trajectory illustrating the time evolution of the molecular copy number $n$ (red solid line), with the blue shaded region denoting the ON state. The trajectory was generated using parameters $\alpha_{\rm on} = 3.0$, $\alpha_{\rm off} = 1.0$, $\beta_{\rm on} = 0.1$, $\beta_{\rm off} = 0.2$, and $k_{\rm on} = k_{\rm off} = 0.01$. 
\textbf{(d)} The probability distribution profile for the time trajectory shown in (c). The simulated joint probability distributions are shown along with the analytical solution (solid, dashed, and dash-dotted lines) for the ON (yellow histogram) and OFF states (light brown histogram), respectively. The inset compares the total probability distribution profile obtained from simulation with the analytical calculation.}
    \label{fig:1}
\end{figure}

\section{RESULTS}
\subsection{Model for generalized accumulation kinetics}
We consider a stochastic model of a genetic switch that alternates between two discrete activity states, denoted by ON and OFF (Fig.~\ref{fig:1}(a)). The downstream molecular species, characterized by copy number $n$, undergoes state-dependent stochastic production and degradation. As illustrated in Fig.~\ref{fig:1}(a), molecules are synthesized at constant rates $\alpha_{\rm on}$ and $\alpha_{\rm off}$ in the ON and OFF states, respectively, while degradation proceeds through first-order kinetics with state-dependent rate constants $\beta_{\rm on}$ and $\beta_{\rm off}$. Stochastic transitions between the two states occur at rates $k_{\rm on}$ and $k_{\rm off}$. To describe the above dynamics systematically, we represent the process in the joint state space $(S,n)$ as a Markov network (Fig.~\ref{fig:1}(b)). The states and the molecular copy number evolve through discrete stochastic transitions. Here $S\in\{\mathrm{ON},\mathrm{OFF}\}$ denotes the state of the system and $n$ is the molecular copy number. The horizontal transitions in the network correspond to synthesis (forward arrows) and degradation (backward arrows) within a given state, whereas vertical transitions represent switching between the two states. The state space shows all possible transitions. The dynamics can form closed loops. The smallest loop connects the four states $(\mathrm{ON},n)\to (\mathrm{ON},n{+}1)\to (\mathrm{OFF},n{+}1)\to (\mathrm{OFF},n)\to (\mathrm{ON},n)$, showing one event for each rate (production, forward switching, degradation and reverse switching). In the absence of vertical transitions, the two states become decoupled and approach their respective local steady states denoted by $\phi_{\rm on}=\alpha_{\rm on}/\beta_{\rm on}$ and $\phi_{\rm off}=\alpha_{\rm off}/\beta_{\rm off}$ for ON and OFF states, respectively. 

The stochastic dynamics of the model, simulated using the Gillespie algorithm \cite{Gillespie1977}, is shown in Fig.~\ref{fig:1}(c). The red line shows the trajectory of copy number $n(t)$. The blue shaded region marks the ON state. The trajectory shows that in each state, the system relaxes towards its local steady state ($\phi_{\rm on}$ and $\phi_{\rm off}$). The timescale of this relaxation dynamics is dictated by degradation rates $\beta_{\rm on}$ and $\beta_{\rm off}$. The corresponding probability distribution of the copy number $n$ in the ON (yellow) and OFF (light brown) states is shown in Fig.~\ref{fig:1}(d). The probability distribution is computed from the fraction of time spent in each state in a given stochastic trajectory. The inset (Fig.~\ref{fig:1}(d)) shows the total probability distribution. For the given parameter set, we obtain a bimodal distribution centered around the characteristic accumulation levels $\phi_{\rm on}$ and $\phi_{\rm off}$.

The state space of the model (Fig.~\ref{fig:1}(b)) can be described by the corresponding chemical master equation (CME) \cite{VanKampen1992}, which governs the time evolution of the joint probability distribution over activity state and copy number. Defining $P_n(t)$ and $Q_n(t)$ as the probabilities of finding the system in the ON and OFF states, respectively, with exactly $n$ molecules at time $t$, the dynamics are governed by
\begin{align}
    \frac{dP_n(t)}{dt} &= \alpha_{\rm on}[P_{n-1}(t)-P_n(t)]
        +\beta_{\rm on}[(n+1)P_{n+1}(t)-nP_n(t)]-k_{\rm off}P_n(t)+k_{\rm on}Q_n(t),  \label{eq:cme_on}\\[2pt]
    \frac{dQ_n(t)}{dt} &= \alpha_{\rm off}[Q_{n-1}(t)-Q_n(t)]
        +\beta_{\rm off}[(n+1)Q_{n+1}(t)-nQ_n(t)]+k_{\rm off}P_n(t)-k_{\rm on}Q_n(t). \label{eq:cme_off}
\end{align}
The first two terms in both equations describe synthesis and first-order degradation within the given state, whereas the remaining terms account for stochastic switching between the ON and OFF states. The total probability is conserved, $\sum_{n=0}^{\infty}\left(P_n (t)+Q_n (t)\right)=1$. Together, these coupled master equations completely specify the time evolution of the system's probability and determine stationary probability distributions at the steady state.

The coupled master equations admit an exact steady-state solution through the probability generating function (PGF) formalism \cite{VanKampen1992,IyerBiswas2009}. We introduce the generating functions $G_{\rm on}(z,t)=\sum_{n=0}^{\infty}P_n(t)\,z^n$ and $G_{\rm off}(z,t)=\sum_{n=0}^{\infty}Q_n(t)\,z^n$, where $z$ is the auxiliary variable with $|z|\leq1$. The conservation of probability requires $G_{\rm on}(1,t)+G_{\rm off}(1,t)=1$. For notational convenience, we define scaled switching rates $\tilde{k}_{\rm off}=k_{\rm off}/\beta_{\rm on}$ and $\tilde{k}_{\rm on}=k_{\rm on}/\beta_{\rm off}$. Substituting the generating functions into the steady-state form of Eqs.~\eqref{eq:cme_on}--\eqref{eq:cme_off} yields a pair of coupled first-order differential equations for $G_{\rm on}(z)$ and $G_{\rm off}(z)$. These can be further written into a single second-order ordinary differential equation, which, after an affine change of variable, assumes the standard form of Kummer's equation. The exact steady-state generating functions are therefore obtained in terms of the confluent hypergeometric (Kummer) function $\mathcal{M}$ \cite{IyerBiswas2009, Ham2020} (see Supplementary Information for the complete derivation),
\begin{align}
    G_{\rm on}(z)  &= \frac{k_{\rm on}}{k_{\rm on}+k_{\rm off}}\,
        \exp\!\left(\phi_{\rm on}(z-1)\right)
        \mathcal{M}\!\left(\tilde{k}_{\rm off},\,1+\tilde{k}_{\rm on}+\tilde{k}_{\rm off};\,
        (\phi_{\rm off}-\phi_{\rm on})(z-1)\right), \label{eq:Gon}\\[3pt]
    G_{\rm off}(z) &= \frac{k_{\rm off}}{k_{\rm on}+k_{\rm off}}\,
        \exp\!\left(\phi_{\rm on}(z-1)\right)
        \mathcal{M}\!\left(1+\tilde{k}_{\rm off},\,1+\tilde{k}_{\rm on}+\tilde{k}_{\rm off};\,
        (\phi_{\rm off}-\phi_{\rm on})(z-1)\right). \label{eq:Goff}
\end{align}
For equal degradation rates, $\beta_{\rm on}=\beta_{\rm off}$, Eqs.~\eqref{eq:Gon}--\eqref{eq:Goff} reduce to the generating function of leaky-telegraph model reported previously \cite{Ham2020,Holehouse2026}. The corresponding stationary distributions follow from the Maclaurin coefficients of Eqs.~\eqref{eq:Gon}--\eqref{eq:Goff},
\begin{align}
    P_n &= \frac{k_{\rm on}}{k_{\rm on}+k_{\rm off}}\,
        \frac{\phi_{\rm on}^{\,n}\,e^{-\phi_{\rm on}}}{n!}
        \sum_{s=0}^{n}\binom{n}{s}
        \left(\frac{\phi_{\rm off}}{\phi_{\rm on}}-1\right)^{\!s}
        \frac{(\tilde{k}_{\rm off})_s}{(1+\tilde{k}_{\rm on}+\tilde{k}_{\rm off})_s}\,
        \mathcal{M}\!\left(\tilde{k}_{\rm off}+s,\,
        1+\tilde{k}_{\rm on}+\tilde{k}_{\rm off}+s;\,
        \phi_{\rm on}-\phi_{\rm off}\right), \label{eq:Pn}\\[3pt]
    Q_n &= \frac{k_{\rm off}}{k_{\rm on}+k_{\rm off}}\,
        \frac{\phi_{\rm off}^{\,n}\,e^{-\phi_{\rm off}}}{n!}
        \sum_{s=0}^{n}\binom{n}{s}
        \left(\frac{\phi_{\rm on}}{\phi_{\rm off}}-1\right)^{\!s}
        \frac{(\tilde{k}_{\rm on})_s}{(1+\tilde{k}_{\rm on}+\tilde{k}_{\rm off})_s}\,
        \mathcal{M}\!\left(\tilde{k}_{\rm on}+s,\,
        1+\tilde{k}_{\rm on}+\tilde{k}_{\rm off}+s;\,
        \phi_{\rm off}-\phi_{\rm on}\right), \label{eq:Qn}
\end{align}
where $(a)_s$ denotes the Pochhammer symbol. The total distribution is estimated as $\psi_n=P_n+Q_n$. The lines (solid, dashed, and dash-dotted lines) in Fig.1(d) show excellent agreement between our analytical distributions and the distributions obtained from stochastic simulations. Taking a sum of Eqs.~\eqref{eq:Pn}--\eqref{eq:Qn} over all $n$ values recovers the stationary occupation probabilities of the two states as $P=k_{\rm on}/(k_{\rm on}+k_{\rm off})$ and $Q=k_{\rm off}/(k_{\rm on}+k_{\rm off})$.  We use the above analytical expression for $\psi_n$ for all subsequent analyses of moments, fluctuations, and thermodynamic quantities. The consistency of the solution can be checked by setting $\phi_{\rm on}=\phi_{\rm off}=\phi$. In this case, the two states become statistically indistinguishable, the Kummer functions reduce to unity, and the total probability collapses to a Poisson distribution, $\psi_n=\phi^{\,n}e^{-\phi}/n!$, i.e., the solution for a classical immigration--death process. 

\begin{figure}[t]
    \centering
    \includegraphics[width=1.0\linewidth]{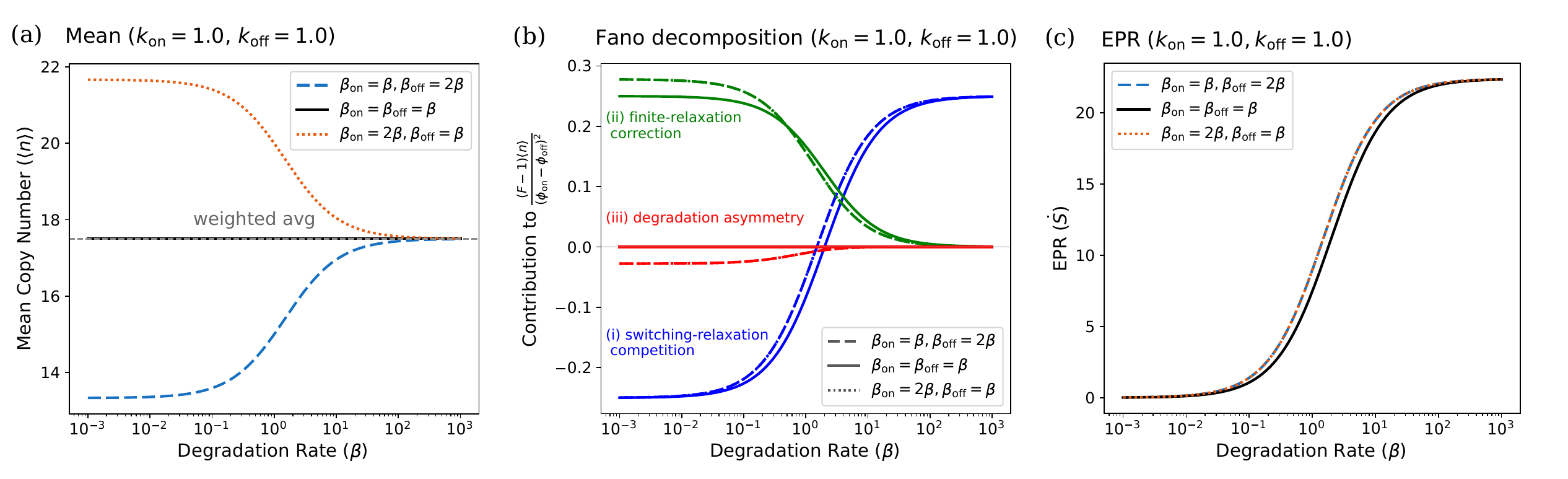}
    \caption{\textbf{Variation of mean, Fano factor components, and EPR with degradation rate.} \textbf{(a)} Variation of mean molecular copy number ($\langle n \rangle$) with the degradation rate ($\beta$). The grey dashed line corresponds to the first term (weighted average) in the mean expression (Eq.~\ref{eq:mean}). \textbf{(b)} Variation of each term in Fano factor expression (Eq.~\ref{eq:fano}) (which contributes to the excess Fano factor, $\frac{(F-1)\langle n \rangle}{(\phi_{\rm on}-\phi_{\rm off})^2}$) with degradation rate $\beta$. The curves (dashed and dotted lines) for two asymmetric cases overlap for each term. \textbf{(c)} Entropy production rate ($\dot{S}$) as a function of $\beta$. The two asymmetric cases (dashed and dotted lines) overlap. All the three graphs are generated using the parameters $k_{\rm on}=k_{\rm off}=1.0$ and $\phi_{\rm on}=30.0$, $\phi_{\rm off}=5.0$. }
    \label{fig:2}
\end{figure}

The exact moments of the steady-state probability distribution can be computed from derivatives of the generating function at $z=1$. The mean copy number ($\langle n\rangle = G'_{\rm on}(1)+G'_{\rm off}(1)$) is given by
\begin{equation}
    \langle n\rangle=
    \frac{k_{\rm on}\phi_{\rm on}+k_{\rm off}\phi_{\rm off}}
        {k_{\rm on}+k_{\rm off}}
    +\frac{k_{\rm off}\tilde{k}_{\rm on}-k_{\rm on}\tilde{k}_{\rm off}}
        {(k_{\rm on}+k_{\rm off})(1+\tilde{k}_{\rm on}+\tilde{k}_{\rm off})}
        (\phi_{\rm on}-\phi_{\rm off}).
    \label{eq:mean}
\end{equation}
The first term in the mean is the occupation-probability-weighted average ($P\phi_{\rm on}+Q\phi_{\rm off})$ of the local steady-state accumulation levels of the ON and OFF states. The second term vanishes in the case of the symmetric-degradation limit ($\beta_{\rm on}=\beta_{\rm off}$ where  $\tilde{k}_{\rm on}/\tilde{k}_{\rm off}=k_{\rm on}/k_{\rm off}$) (see Fig.~\ref{fig:2}(a)) and recovers previous results \cite{Ham2020, Holehouse2026}. Thus, in the case of state-dependent degradation rates, the mean value is modulated by the difference in the local steady-state levels and the switching rates between the states. To study how the degradation rates modulate the mean copy number, we study the variation of $\langle n \rangle$ for fixed values of $\phi_{\rm on}$ and $\phi_{\rm off}$ and equal switching rates in Fig. 2(a). The degradation dependent term either increases or decreases the mean from $\langle n \rangle \approx \frac{\phi_{\rm on}+\phi_{\rm off}}{2}$ (the symmetric case, $\beta_{\rm on}=\beta_{\rm off}$, black line) depending on if $\beta_{\rm on}>\beta_{\rm off}$ or $\beta_{\rm on}<\beta_{\rm off}$. In contrast, in the fast-switching limit  ($k_{\rm on},k_{\rm off}\gg \beta_{\rm on},\beta_{\rm off}$), the state of the system changes rapidly and the mean approaches to an intermediate level, the switching-averaged value $\langle n \rangle \to \frac{\phi_{\rm on}\beta_{\rm on}+\phi_{\rm off}\beta_{\rm off}}{\beta_{\rm on}+\beta_{\rm off}}$. 
The variance of copy number can be analogously estimated from the second derivative of the generating function at $z=1$; owing to its length, the explicit expression is given in the Supplementary Information.

\subsection{Fano factor as a measure of statistical fluctuation}
\label{sec:fano}

To characterize the fluctuations in the steady-state with respect to the mean expression, we use the Fano factor $F=\sigma^2/\langle n\rangle$, where $\sigma^2$ is the variance \cite{banerjee2025fano}. Smaller $F$ signifies higher statistical precision, and larger $F$ enhances the variability. Fano factor, $F=1$, represents a Poisson distribution where the mean and variance are the same. Substituting the exact first two moments (Supplementary Information) gives
\begin{eqnarray}
F=1+\frac{(\phi_{\rm on}-\phi_{\rm off})^2}
        {\langle n\rangle}
    \Bigg[
        \frac{k_{\rm on}k_{\rm off}\,(1-\tilde{k}_{\rm on}-\tilde{k}_{\rm off})}
            {(k_{\rm on}+k_{\rm off})^2(1+\tilde{k}_{\rm on}+\tilde{k}_{\rm off})}
        +\frac{k_{\rm on}\tilde{k}_{\rm off}(1+\tilde{k}_{\rm off})
            +k_{\rm off}\tilde{k}_{\rm on}(1+\tilde{k}_{\rm on})}
            {(k_{\rm on}+k_{\rm off})(1+\tilde{k}_{\rm on}+\tilde{k}_{\rm off})(2+\tilde{k}_{\rm on}+\tilde{k}_{\rm off})}\nonumber
       \\ -\frac{(k_{\rm on}\tilde{k}_{\rm off}-k_{\rm off}\tilde{k}_{\rm on})^2}
            {(k_{\rm on}+k_{\rm off})^2(1+\tilde{k}_{\rm on}+\tilde{k}_{\rm off})^2}\Bigg].
    \label{eq:fano}
\end{eqnarray}
The deviation of the fluctuations from the classical immigration-death process ($F=1$) is directly modulated by the separation of the local steady-state accumulations through the term $(\phi_{\rm on}-\phi_{\rm off})^2$. For identical states, $\phi_{\rm on}=\phi_{\rm off}$, the Fano factor reduces to $1$ as expected for Poissonian statistics. When $k_{\rm on}=0$ or $k_{\rm off}=0$, the system again behaves like a single-state system, resulting in $F=1$. The departure from Poissonian statistics ($F=1$) can be associated with three distinct mechanisms. The first contribution, reflecting the competition between switching and molecular relaxation, enhances fluctuations when $\tilde{k}_{\rm on}+\tilde{k}_{\rm off}<1$ and suppresses them when $\tilde{k}_{\rm on}+\tilde{k}_{\rm off}>1$, (Fig.~\ref{fig:2}(b), blue line). The second contribution can be associated with a finite relaxation time, as it vanishes in the limit of the fast-degradation rate (when $\beta_{\rm on},\beta_{\rm off}\gg k_{\rm on},k_{\rm off}$, i.e.\ $\tilde{k}_{\rm on},\tilde{k}_{\rm off}\ll1$) (Fig.~\ref{fig:2}(b), green line). This term always enhances fluctuations and arises from the residual memory of its previous activity state after a switching event. The final term is associated with the asymmetry in degradation rates, as the numerator vanishes for symmetric degradation (Fig.~\ref{fig:2}(b), red line). Further for $k_{\rm on}\approx k_{\rm off}$, in the fast switching regime ($k_{\rm on},k_{\rm off}\gg\beta_{\rm on},\beta_{\rm off}$), the three terms cancel each other in eq.~\ref{eq:fano} and $F$ approaches 1, showing a near-Poissonian distribution. This condition mimics an effective single steady-state achieved by effective immigration-death rates. In the slow switching regime ($k_{\rm on},k_{\rm off}<<\beta_{\rm on},\beta_{\rm off}$), the second and third terms become negligible, yielding $F\to 1+\frac{(\phi_{\rm on}-\phi_{\rm off})^2}{4\langle n\rangle}$ (Fig.~\ref{fig:2}(b)).

\subsection{Entropy production rate as a measure of thermodynamic cost}
\label{sec:epr}

The coupled dynamics of phenotypic switching and molecular accumulation drive the system into a non-equilibrium steady-state. In the joint state space ($S,n$) (see Fig.~\ref{fig:1} (b)), the stochastic transitions between phenotypic states and molecular copy numbers form closed cycles. At steady-state, the probability distribution becomes time-independent; however, individual transitions continue to occur, allowing probability to circulate through these cycles. When the two phenotypic states have distinct molecular dynamics, such that $\phi_{\rm on}\ne \phi_{\rm off}$, the forward and reverse products of transition rates around a cycle generally differ (see Supplementary Information), violating the detailed-balance condition. Consequently, nonzero probability currents persist in the steady-state, distinguishing the system from thermodynamic equilibrium. To quantify the thermodynamic cost of maintaining this non-equilibrium state, we evaluate the entropy production rate (EPR). We use Schnakenberg's cycle decomposition, where the EPR is written as a sum over the independent kinetic cycles \cite{Schnakenberg1976},
\begin{equation}
    \dot{S} = k_B\sum_{\mathrm{cycle}}J_{\mathrm{cycle}}\,\mathcal{A}_{\mathrm{cycle}},
    \label{eq:schnakenberg}
\end{equation}
with $J_{\mathrm{cycle}}$ the current circulating around an elementary cycle and $\mathcal{A}_{\mathrm{cycle}}$ its thermodynamic affinity. For our model, the translational symmetry of the network along the copy-number axis ensures that every elementary cycle carries the same affinity. For the plaquette $(\mathrm{ON},n)\to(\mathrm{ON},n{+}1)\to(\mathrm{OFF},n{+}1)\to(\mathrm{OFF},n)\to
(\mathrm{ON},n)$, the affinity is the log-ratio of the forward ($\alpha_{\rm on} k_{\rm off}\beta_{\rm off}(n+1) k_{\rm on}$) and reverse ($\alpha_{\rm off} k_{\rm on}\beta_{\rm on}(n+1) k_{\rm off}$) rate products,
\begin{equation}
    \mathcal{A}=\ln\!\left(\frac{\phi_{\rm on}}{\phi_{\rm off}}\right)=\ln\!\left(\frac{\alpha_{\rm on}}{\alpha_{\rm off}}\right)
    +\ln\!\left(\frac{\beta_{\rm off}}{\beta_{\rm on}}\right)
    \label{eq:affinity_split}
\end{equation}
where the factors $((n+1),k_{\rm on},k_{\rm off})$ cancel between the forward and reverse directions, leaving an affinity independent of state $n$ and switching rates (See Supplementary Information). The affinity can be further decomposed into contributions from relative change in accumulation and degradation in ON and OFF states (Eq.~\ref{eq:affinity_split}). Similarly, evaluating the stationary currents from the exact solution gives the total cycle current as (see Supplementary Information for detailed derivation),
\begin{equation}
    \sum_{\mathrm{cycle}}J_{\mathrm{cycle}}
    =\frac{k_{\rm on}k_{\rm off}}
        {(k_{\rm on}+k_{\rm off})(1+\tilde{k}_{\rm on}+\tilde{k}_{\rm off})}
        (\phi_{\rm on}-\phi_{\rm off}),
    \label{eq:cycle_current}
\end{equation}
and hence, setting $k_B=1$, the exact steady-state entropy production rate (EPR) is
\begin{equation}
    \dot{S}=\frac{k_{\rm on}k_{\rm off}}
        {(k_{\rm on}+k_{\rm off})(1+\tilde{k}_{\rm on}+\tilde{k}_{\rm off})}
        (\phi_{\rm on}-\phi_{\rm off})\,
        \ln\!\left(\frac{\phi_{\rm on}}{\phi_{\rm off}}\right)
    \label{eq:epr}
\end{equation}
The EPR is the product of the cyclic current and the affinity, and is always positive. It vanishes when two states are identical, $\phi_{\rm on}=\phi_{\rm off}$, where both the affinity and the circulating current are zero. The EPR is also zero for either $k_{\rm on}=0$ or $k_{\rm off}=0$, as the system has only one state and it satisfies detailed balance. Fig.~\ref{fig:2}(c) shows the variation of the EPR with degradation rate for fixed switching rates ($k_{\rm on}= k_{\rm off}=k$). The EPR curve increases from a low level and saturates to a finite value as degradation increases. For small degradation rates ($\beta_{\rm on},\beta_{\rm off}\ll k$), the EPR is proportional to the degradation rate ($\dot{S}\to \frac{\beta}{4} (\phi_{\rm on}-\phi_{\rm off})\ln\left(\phi_{\rm on}/\phi_{\rm off}\right)$ for $\beta_{\rm on}=\beta_{\rm off}=\beta$), and for large degradation rates ($\beta_{\rm on},\beta_{\rm off}\gg k$), the EPR saturates proportional to the switching rate $k$ ($\dot{S}\to\frac{k}{2}(\phi_{\rm on}-\phi_{\rm off})\ln\left(\phi_{\rm on}/\phi_{\rm off}\right)$). For unequal degradation rates, the saturation level remains unchanged, but the EPR slightly increases at intermediate values because the denominator is suppressed as one of the degradation rates is higher ($2\beta$ in this case). This effect is symmetric with respect to the exchange of degradation rates due to the symmetric nature of $\beta$ terms in the denominator for $k_{\rm on}= k_{\rm off}=k$.

\section{State distributions and transitions}
\begin{figure}[t]
    \centering
    \includegraphics[width=1.0\linewidth]{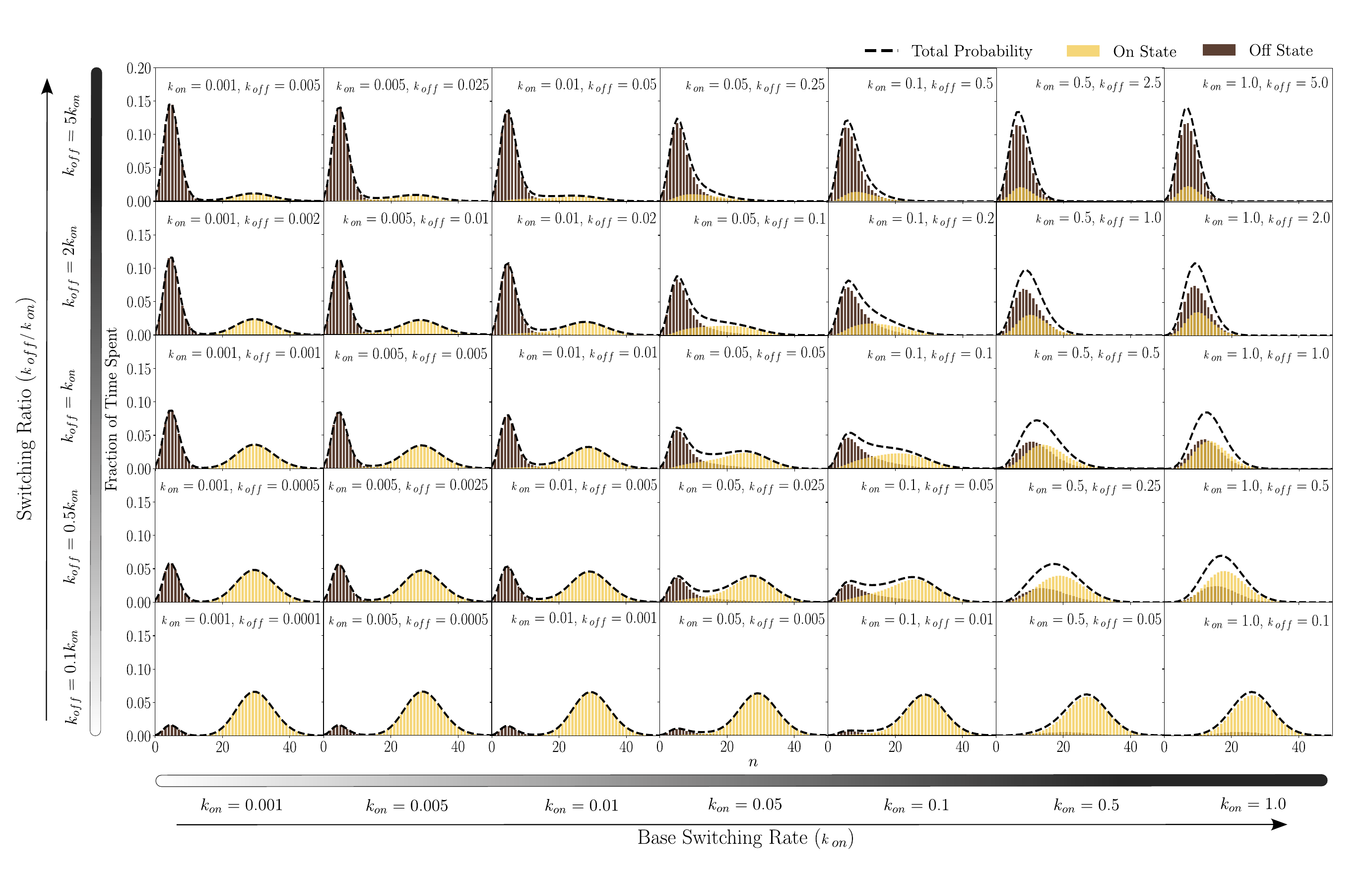}
    \caption{\textbf{Steady-state probability distributions across the different switching rates.} Each panel shows the exact analytical steady-state distributions for the ON state ($P_n$, yellow), OFF state ($Q_n$, brown), and the total probability ($\psi_n = P_n + Q_n$, black dashed curve), computed from Eqs.~\ref{eq:Pn}--\ref{eq:Qn}. The base switching rate $k_{\rm on}$ increases along the horizontal axis (left to right), and the switching bias $k_{\rm off}/k_{\rm on}$ increases along the vertical axis (bottom to top). Parameters are fixed at $\alpha_{\rm on}=3.0$, $\alpha_{\rm off}=1.0$, $\beta_{\rm on}=0.1$, $\beta_{\rm off}=0.2$, giving intrinsic levels $\phi_{\rm on}=30.0$ and $\phi_{\rm off}=5.0$.}
    \label{fig:3}
\end{figure}
To associate the Fano factor and EPR values with the occupation probability of the different accessible states in the state space, we plot the probability distribution (using Eqs. ~\ref{eq:Pn}--\ref{eq:Qn}) by varying the strength and asymmetry of the switching rates between the two states. Motivated by recent experiments by Hung et al.~\cite{Hung2014}, which demonstrated diverse phenotypic distributions in synthetic gene networks by varying the base switching rate $k_{\rm on}$ and the switching ratio $k_{\rm off}/k_{\rm on}$, we construct a phase diagram showing all possible distributions and the transitions between them. Fig.~\ref{fig:3} shows the probability distributions for $n$ as a function of the strength or frequency of the base switching rate $k_{\rm on}$ along the x-axis for a given bias or switching ratio $k_{\rm off}/k_{\rm on}$ along the y-axis.

The frequency of switching controls the transitions between different modalities, i.e., from bimodal to unimodal distribution, as shown in the plots along the rows in Fig.~\ref{fig:3}. When the switching rates are slow relative to degradation rates ($k_{\rm on},k_{\rm off}\ll\beta_{\rm on},\beta_{\rm off}$), the system fully relaxes to its local steady-state level between the transitions. Consequently, the total probability distribution separates into two well-resolved subpopulations centered around $\phi_{\rm
on}$ and $\phi_{\rm off}$. As the frequency of switching increases ($k_{\rm on},k_{\rm off}\gg\beta_{\rm on},\beta_{\rm off}$), the system is repeatedly interrupted before approaching its local steady-state between transitions. Hence, the system resides at the time-averaged level of the two states, and the sub-distributions merge into a single near-Poissonian distribution with a peak at intermediate values.

The effect of a change in the bias of the switching rates is shown along the columns in Fig.~\ref{fig:3}. The bias determines which subpopulation dominates. At slow to intermediate switching frequencies, increasing the bias from $k_{\rm off}/k_{\rm on}\ll1$ (ON-dominated) toward $k_{\rm off}/k_{\rm on}\gg1$ (OFF-dominated) drives the system from a high-mean unimodal distribution to a low-mean unimodal distribution through an intermediate bimodal distribution. The peak height corresponding to the ON state decreases progressively while that for the OFF state increases, while the location of both peaks remains fixed at $\phi_{\rm on}$ and $\phi_{\rm off}$. In contrast, at high frequency of switching, the system transitions directly from an ON-dominated to an OFF-dominated distribution without passing through a bimodal intermediate. In this regime, the peak of the unimodal distribution shifts continuously from $\phi_{\rm on}$ to $\phi_{\rm off}$, thereby changing the mean without altering the modality. Together, these paths identify all possible transitions (realized in a synthetic gene network  \cite{Hung2014}) in the distribution as the frequency and bias of the switching rates between the states are tuned in the model.
To quantify the uncertainty of observed states across this distribution space, we also computed the Shannon entropy $H=-\sum_n\psi_n\ln\psi_n$ using the exact distributions (see Supplementary Information). The entropy is maximal at intermediate switching rates ($k_{\rm on}\approx 0.056$, $k_{\rm off}\approx 0.027$), where the two sub-distributions are merging, and probability is spread broadly across copy-number states. In contrast, the entropy is minimal at the boundaries of the distribution space, i.e., at extreme switching bias ratios and at very fast switching, where the joint distribution reduces to a narrow unimodal profile.

\subsection{Fano factor and EPR variation during transitions}
To further characterize how statistical fluctuation and thermodynamic dissipation vary during these transitions, we analyze the variation of the Fano factor and the entropy production rate during the shift in distributions. We study the variation of the Fano factor and the EPR with mean copy number, which has been used to characterize different networks in previous studies \cite{zhang2022rate, Holehouse2026, banerjee2025fano}. It is known that for a system with only one state (either ON or OFF), the molecular concentration displays a Poisson distribution with $F=1$ and transitions between the neighboring states obey detailed balance, leading to zero thermodynamic cost ($\dot{S}=0$). 

\begin{figure}[t]
    \centering
\includegraphics[width=1.0\linewidth]{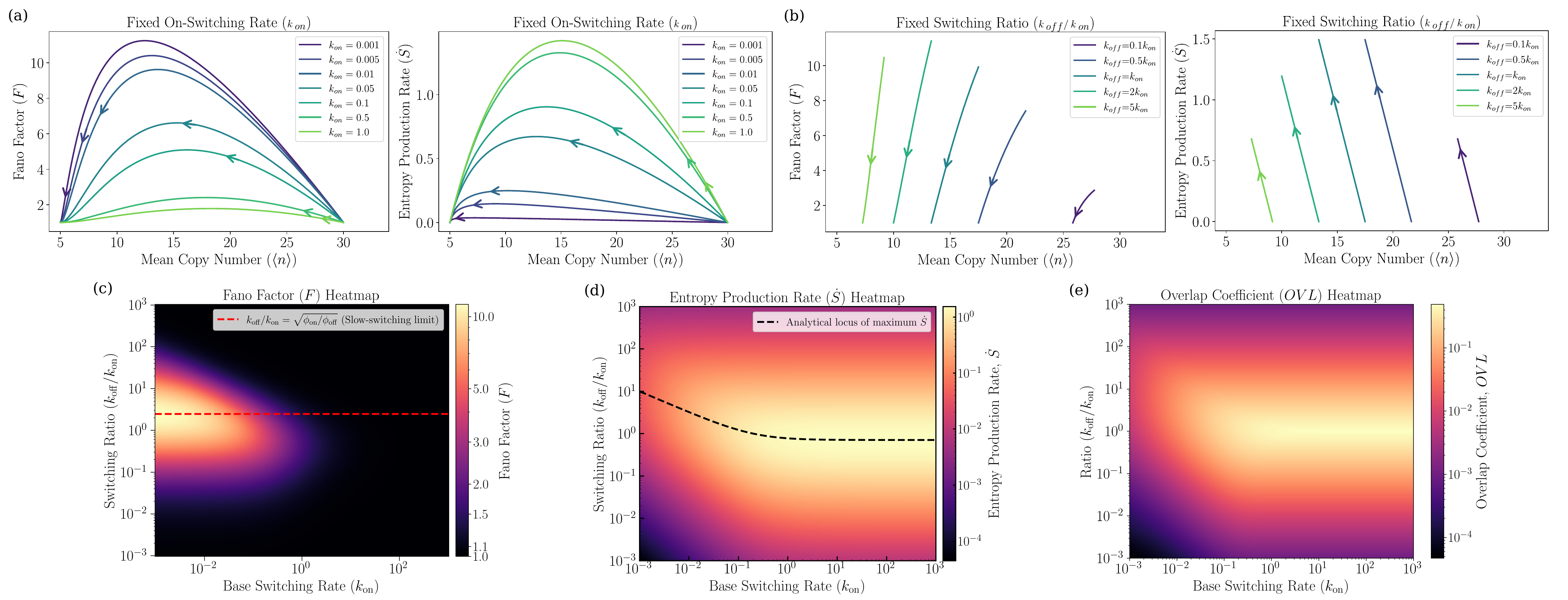}
    \caption{\textbf{Variation of Fano factor and EPR.} \textbf{(a)} Fano factor $F$ (left sub-panel) and entropy production rate $\dot{S}$ (right sub-panel) plotted against the mean copy number $\langle n \rangle$ as parametric curves, with the base switching rate $k_{\rm on}$ fixed and the bias $k_{\rm off}/k_{\rm on}$ is varied (arrows indicate the direction of increasing $k_{\rm off}$). Each curve traces a characteristic curve: at extreme bias one state dominates, the distribution is near-Poissonian, and both $F \approx1$ and $\dot{S}\approx0$; at intermediate bias both states are substantially occupied, $F$ and $\dot{S}$ reach their respective maxima. \textbf{(b)} Fano factor $F$ (left sub-panel) and EPR $\dot{S}$ (right sub-panel) plotted with the switching ratio $k_{\rm off}/k_{\rm on}$  fixed and both $k_{\rm on}$ and $k_{\rm off}$ are varied (arrows indicate the direction of increasing $k_{\rm on},k_{\rm off}$). The Fano factor decreases monotonically as switching frequency rises, while EPR increases. The arrows in the two sub-panels point in opposite directions along the $\langle n\rangle$ axis. \textbf{(c)} Fano factor heatmap over the base switching rate $k_{\rm on}$ (horizontal) and switching ratio $k_{\rm off}/k_{\rm on}$ (vertical). The red dashed line marks the analytical locus of maximum Fano factor in the slow-switching limit (Eq.~\ref{eq:fano_locus}). \textbf{(d)} Entropy production rate heatmap. The largest dissipation (brightest region) is observed in the fast-switching limit and moderate bias. The black dashed curve marks the exact analytical locus of maximal dissipation (Eq.~\ref{eq:epr_locus} ). \textbf{(e)} Overlap between the subpopulation distributions. The overlap is large in the fast-switching limit where two sub-distributions merge.}
    \label{fig:4}
\end{figure}

Fig.~\ref{fig:4}(a) shows the Fano factor $F$ (left sub-panel) and entropy production rate $\dot{S}$ (right sub-panel) as functions of the mean copy number $\langle n \rangle$, for fixed base switching rate $k_{\rm on}$ and varying switching ratio $k_{\rm off}/k_{\rm on}$, corresponding to traversing a column of Fig.~\ref{fig:3}. Arrows indicate the direction of increasing $k_{\rm off}$. At low $k_{\rm off}$, the ON state dominates, the mean is close to $\phi_{\rm on}$, and the distribution is near-Poissonian, giving $F\approx1$ and $\dot{S}\approx0$. As $k_{\rm off}$ increases, both states become substantially occupied; the resulting distribution drives $F$ and $\dot{S}$ to increase toward their respective maxima at intermediate $k_{\rm off}$. As $k_{\rm off}$ increases further, the OFF state dominates, the distribution is near-Poissonian around $\phi_{\rm off}$, and both $F$ and $\dot{S}$ return to one and zero, respectively. Every curve therefore traces a characteristic path, beginning and ending near $F\approx1$ ($\dot{S}\approx0$) as the mean shifts from $\phi_{\rm on}$ to $\phi_{\rm off}$. 

Further, as $k_{\rm on}$ increases, the mean at which both $F$ and $\dot{S}$ reach their respective maxima shifts to higher values, while their overall values show an inverse relationship—one decreasing while the other increases with $k_{\rm on}$. For small $k_{\rm on}$ values, when the switching bias is increased, the distribution transitions from a high-mean unimodal, through a bimodal intermediate, to a low-mean unimodal profile. In contrast, for fast switching, the distribution shifts directly from a high-mean to a low-mean unimodal profile without passing through a bimodal intermediate. Hence, $F$ reaches higher values for small $k_{\rm on}$ due to the presence of the intermediate bimodal distribution, and lower values at high $k_{\rm on}$ due to the merged unimodal distribution, which remains close to Poisson. Consequently, because switching events are rare (for small $k_{\rm on}$ values), $\dot{S}$ remains low as few probability cycles complete per unit time. Conversely, when switching is fast, $\dot{S}$ attains higher values because the number of probability cycles per unit time is large. A similar trend is seen in Fig.~\ref{fig:4}(b), where both $k_{\rm on}$ and $k_{\rm off}$ are varied simultaneously for a fixed switching rate ratio $k_{\rm off}/k_{\rm on}$ (bias), i.e., along the row of Fig.~\ref{fig:3}. The arrows indicate the direction of increase in the switching rates. As expected, the Fano factor (left sub-panel) decreases, and the EPR ($\dot{S}$) increases monotonically as the switching rate increases.

Next, we plot the Fano factor and EPR (Fig. \ref{fig:4} (c), (d)) over the full phenotypic distribution space of Fig.~\ref{fig:3}. Similar to parametric curves plotted before, the Fano-factor heatmap (Fig.~\ref{fig:4}(c)) shows that the highest Fano factor values are in the slow-switching and intermediate-bias regions, where the system is bimodal. The location of the peak can be identified for slow switching rates($k_{\rm on},k_{\rm off}\ll\beta_{\rm on},\beta_{\rm off}$), where the ratio that maximizes $F$ is given by
\begin{equation}
    {\frac{k_{\rm off}}{k_{\rm on}}}^{*}\approx\sqrt{\frac{\phi_{\rm on}}{\phi_{\rm off}}}.
    \label{eq:fano_locus}
\end{equation}
Thus, the optimal ratio is the square root of the ratio of the local steady-state levels (derivation in the Supplementary Information). This asymptote is marked with a red dashed line in Fig.~\ref{fig:4}(c). At the optimal bias, $F\to 1 + (\sqrt{\phi_{\rm on}}-\sqrt{\phi_{\rm off}})^2$, which is the maximum value at the phase space. As switching increases, rapid transitions between states average the distribution toward a unimodal distribution, driving $F\to1$. 

The EPR map (Fig.~\ref{fig:4}(d)) shows opposite variation. The entropy production rate is highest in the fast-switching regime, as the system dissipates maximum energy by rapidly moving from one activity state to another. The curve of maximal dissipation is exactly found and shows the following relation with bias,
\begin{equation}
    {\frac{k_{\rm off}}{k_{\rm on}}}^{*}=\sqrt{\frac{\beta_{\rm on}}{\beta_{\rm off}}
        \left(1+\frac{\beta_{\rm off}}{k_{\rm on}}\right)}
    \label{eq:epr_locus}
\end{equation}
For slower switching rates, the optimal ratio decreases with $k_{\rm on}$ as ${\frac{k_{\rm off}}{k_{\rm on}}}^*\to\sqrt{\frac{\beta_{\rm on}}{k_{\rm on}}}$. Whereas for fast switching rates, it becomes independent of switching rates as ${\frac{k_{\rm off}}{k_{\rm on}}}^*\to\sqrt{\frac{\beta_{\rm on}}{\beta_{\rm off}}}$. The ratio of degradation rates determines the maximum location. At fast-switching rate, we have the maximum dissipation, $\dot{S}\to \frac{\beta_{\rm on}\beta_{\rm off}}{(\sqrt{\beta_{\rm on}}+\sqrt{\beta_{\rm off}})^2}(\phi_{\rm on}-\phi_{\rm off})ln\left(\frac{\phi_{\rm on}}{\phi_{\rm off}}\right)$. Finally to measure the extent of the geometric overlap between the two subpopulations, we compute the overlap coefficient, using the analytical expression of  $P_n$ and $Q_n$, as $\mathrm{OVL}=\sum_{n=0}^{\infty}\min(P_n,Q_n)$ which measures the shared probability mass of the two states (Fig.~\ref{fig:4}(e)). The overlap is bounded by $0\le\min(P,Q)\le\tfrac12$. The upper bound of $\tfrac12$ is approached only under complete mixing of an unbiased system ($k_{\rm on}\approx k_{\rm off}$, fast switching), where $P_n$ and $Q_n$ acquire the same shape and become statistically indistinguishable. The overlap vanishes both when the peaks are well separated (slow switching) and when one state is strongly suppressed (extreme bias). The map shows that the regime of maximal dissipation (large EPR values) corresponds to the maximum overlap in subpopulation distributions.

\section{Discussion}

Stochastic switching between different gene expression states dictates the accumulation of mRNA, proteins, and other molecules in cells \cite{elowitz2002stochastic, Karmakar2020, Arkin1997, kaern2005stochasticity}. While the mechanisms underlying bistability and noise-induced switching are extensively characterized, far less is understood about how distributions of downstream entities vary as kinetic parameters such as switching rates, production rates, and degradation rates change \cite{de2020shape, shahrezaei2008analytical, friedman2006linking}. To fill this knowledge gap, we generalize the two-state stochastic model for mRNA accumulation dynamics to include state-dependent production and degradation rates. Using this model, we compute fully analytical expressions for the concentration distribution and associated quantitative metrics, such as the mean, Fano factor, and entropy production rate. By simplifying these expressions, we dissect the individual contributions of each kinetic parameter to these quantities. We find that state-dependent degradation rates introduce an additional term to the mean, extending beyond the standard occupation-probability-weighted average. While previous studies of mRNA dynamics typically assume identical degradation rates, we show that even slight deviations can get amplified to cause a significant shift in the mean. 

Using the analytical expressions for the probability distributions, we map all possible modalities and transitions between them as switching rates are varied.  The variation of the distribution across switching rates identifies distinct transition types: a transition from/to a high-mean unimodal to/from a low-mean state through a bimodal intermediate, a transition from/to a bimodal to/from a fused unimodal state with an intermediate mean, and a shift in the fused unimodal distribution from/to high-mean to/from low-mean without passing through a bimodal state. Experimentally, similar transitions between different modalities have been observed in synthetic genetic networks through the modulation of kinetic parameters \cite{becskei2001positive, Hung2014}. Furthermore, such shifts in RNA and protein concentration distributions occur through the use of inducible systems and regulatory factor titration \cite{palme2021variation, becskei2001positive, portle2007cell}, across distinct cell cycle phases \cite{mcdavid2014modeling}, via genetic mutations \cite{mason2011bimodal, baptista2025bimodality}, and during environmental shifts \cite{spratt2026temporal}.

In summary, our model and analytical probability distributions can be used to fit experimental data and extract parameters, serve as a null model to assess the impact of differential mRNA stability across transcriptional states, and simulate other processes where stochastic switching is coupled to downstream dynamics—such as two-state protein dynamics (with fast mRNA kinetics) and drug accumulation. Furthermore, these findings illuminate the fundamental physics governing non-equilibrium kinetics and phenotypic distributions in biological networks, thereby offering new design principles for synthetic systems.

\bibliography{References}

@article{Elowitz2002,
	author = {Elowitz, Michael B and Levine, Arnold J and Siggia, Eric D and Swain, Peter S},
	doi = {10.1126/science.1070919},
	journal = {{Science}},
	month = aug,
	nlmuniqueid = {0404511},
	number = {5584},
	pages = {1183--6},
	pii = {297/5584/1183},
	pubmed = {12183631},
	title = {{Stochastic gene expression in a single cell.}},
	volume = {297},
	x-fetchedfrom = {PubMed},
	year = {2002}
}

@article{Raj2008,
	author = {Raj, A. and {van Oudenaarden}, A.},
	journal = {Cell},
	number = 2,
	pages = {216--226},
	publisher = {Elsevier},
	title = {{Nature, nurture, or chance: stochastic gene expression and its consequences}},
	url = {http://www.bibsonomy.org/bibtex/23328343571931ad09f5795856274f243/peter.ralph},
	volume = 135,
	x-fetchedfrom = {Bibsonomy},
	year = 2008
}

@article{Altschuler2010,
  author  = {Altschuler, Steven J. and Wu, Lani F.},
  title   = {Cellular heterogeneity: do differences make a difference?},
  journal = {Cell},
  volume  = {141}, number = {4}, pages = {559--563}, year = {2010},
  publisher = {Elsevier}
}

@article{kaern2005stochasticity,
  title={Stochasticity in gene expression: from theories to phenotypes},
  author={Kaern, Mads and Elston, Timothy C and Blake, William J and Collins, James J},
  journal={Nature Reviews Genetics},
  volume={6},
  number={6},
  pages={451--464},
  year={2005},
  publisher={Nature Publishing Group UK London}
}

@article{Balaban2004,
	author = {Balaban, Nathalie Q. and Merrin, Jack and Chait, Remy and Kowalik, Lukasz and Leibler, Stanislas},
	description = {persister growth and switching rates},
	doi = {10.1126/science.1099390},
	journal = {Science},
	number = 5690,
	pages = {1622--1625},
	title = {{Bacterial Persistence as a Phenotypic Switch}},
	url = {http://www.sciencemag.org/content/305/5690/1622.abstract},
	username = {quantentunnel},
	volume = 305,
	x-fetchedfrom = {Bibsonomy},
	year = 2004
}

@article{Kussell2005,
  author  = {Kussell, Edo and Leibler, Stanislas},
  title   = {Phenotypic diversity, population growth, and information in
             fluctuating environments},
  journal = {Science},
  volume  = {309}, number = {5743}, pages = {2075--2078}, year = {2005},
  publisher = {AAAS}
}

@article{Paulsson2004,
  author  = {Paulsson, Johan},
  title   = {Summing up the noise in gene networks},
  journal = {Nature},
  volume  = {427}, number = {6973}, pages = {415--418}, year = {2004},
  publisher = {Nature Publishing Group}
}

@article{Karmakar2020,
  author  = {Karmakar, Ranadhir and Dey, Sandip},
  title   = {Stochastic model of gene expression with slow and fast
             switching},
  journal = {Physical Review E},
  volume  = {102}, number = {4}, pages = {042401}, year = {2020},
  publisher = {American Physical Society}
}

@article{eldar2010functional,
  title={Functional roles for noise in genetic circuits},
  author={Eldar, Avigdor and Elowitz, Michael B},
  journal={Nature},
  volume={467},
  number={7312},
  pages={167--173},
  year={2010},
  publisher={Nature Publishing Group UK London}
}

@article{munsky2012using,
  title={Using gene expression noise to understand gene regulation},
  author={Munsky, Brian and Neuert, Gregor and Van Oudenaarden, Alexander},
  journal={Science},
  volume={336},
  number={6078},
  pages={183--187},
  year={2012},
  publisher={American Association for the Advancement of Science}
}

@article{Swain2002,
  author  = {Swain, Peter S. and Elowitz, Michael B. and Siggia, Eric D.},
  title   = {Intrinsic and extrinsic contributions to stochasticity in
             gene expression},
  journal = {Proceedings of the National Academy of Sciences},
  volume  = {99}, number = {20}, pages = {12795--12800}, year = {2002},
  publisher = {National Academy of Sciences}
}

@article{Raj2006,
  author  = {Raj, Arjun and Peskin, Charles S. and Tranchina, Daniel and
             Vargas, Diana Y. and Tyagi, Sanjay},
  title   = {Stochastic {mRNA} synthesis in mammalian cells},
  journal = {PLoS Biology},
  volume  = {4},
  number  = {10},
  pages   = {e309},
  year    = {2006},
  publisher = {Public Library of Science}
}

@article{Golding2005,
  author  = {Golding, Ido and Paulsson, Johan and Zawilski, Scott M.
             and Cox, Edward C.},
  title   = {Real-time kinetics of gene activity in individual bacteria},
  journal = {Cell},
  volume  = {123}, number = {6}, pages = {1025--1036}, year = {2005},
  publisher = {Elsevier}
}

@article{Patra2013,
	author = {Patra, Pintu and Klumpp, Stefan},
	doi = {10.1371/journal.pone.0062814},
	journal = {{PLoS ONE}},
	nlmuniqueid = {101285081},
	number = {5},
	pages = {e62814},
	pii = {PONE-D-12-36324},
	pmc = {PMC3652822},
	pubmed = {23675428},
	title = {{Population dynamics of bacterial persistence.}},
	volume = {8},
	x-fetchedfrom = {PubMed},
	year = {2013}
}

@article{Losick2008,
  author  = {Losick, Richard and Desplan, Claude},
  title   = {Stochasticity and cell fate},
  journal = {Science},
  volume  = {320}, number = {5872}, pages = {65--68}, year = {2008},
  publisher = {AAAS}
}

@article{Gupta2011,
  author  = {Gupta, Piyush B. and Fillmore, Christine M. and Jiang, Guozhi
             and Shapira, Sagi D. and Tao, Kai and Kuperwasser, Charlotte
             and Lander, Eric S.},
  title   = {Stochastic state transitions give rise to phenotypic equilibrium
             in populations of cancer cells},
  journal = {Cell},
  volume  = {146}, number = {4}, pages = {633--644}, year = {2011},
  publisher = {Elsevier}
}

@article{Lu2013,
  author  = {Lu, Mingyang and Jolly, Mohit Kumar and Levine, Herbert and
             Onuchic, Jos\'{e} N. and Ben-Jacob, Eshel},
  title   = {MicroRNA-based regulation of epithelial--hybrid--mesenchymal
             fate determination},
  journal = {Proceedings of the National Academy of Sciences},
  volume  = {110}, number = {45}, pages = {18144--18149}, year = {2013},
  publisher = {National Academy of Sciences}
}

@article{Acar2008,
  author  = {Acar, Murat and Mettetal, Jerome T. and van Oudenaarden,
             Alexander},
  title   = {Stochastic switching as a survival strategy in fluctuating
             environments},
  journal = {Nature Genetics},
  volume  = {40}, number = {4}, pages = {471--475}, year = {2008},
  publisher = {Nature Publishing Group}
}

@article{Raser2005,
  author  = {Raser, Jonathan M. and O'Shea, Erin K.},
  title   = {Noise in gene expression: origins, consequences, and control},
  journal = {Science},
  volume  = {309}, number = {5743}, pages = {2010--2013}, year = {2005},
  publisher = {AAAS}
}

@article{Hung2014,
  author  = {Hung, Michelle and Chang, Emily and Hussein, Razika and Frazier, Katya
             and Shin, Jung-Eun and Sagawa, Shiori and Lim, Han N.},
  title   = {Modulating the frequency and bias of stochastic switching to control
             phenotypic variation},
  journal = {Nature Communications},
  volume  = {5},
  pages   = {4574},
  year    = {2014},
  publisher = {Nature Publishing Group},
  doi     = {10.1038/ncomms5574}
}

@article{Peccoud1995,
  author  = {Peccoud, Jean and Ycart, Bernard},
  title   = {Markovian modeling of gene-product synthesis},
  journal = {Theoretical Population Biology},
  volume  = {48},
  number  = {2},
  pages   = {222--234},
  year    = {1995},
  publisher = {Elsevier}
}

@article{IyerBiswas2009,
  author  = {Iyer-Biswas, Srividya and Hayot, F. and Jayaprakash, C.},
  title   = {Stochasticity of gene products from transcriptional pulsing},
  journal = {Physical Review E},
  volume  = {79},
  number  = {3},
  pages   = {031911},
  year    = {2009},
  publisher = {American Physical Society}
}

@article{Arkin1997,
	author = {McAdams, Harley H and Arkin, Adam},
	journal = {Proceedings of the National Academy of Sciences},
	number = {3},
	pages = {814--819},
	publisher = {National Acad Sciences},
	title = {{Stochastic mechanisms in gene expression}},
	url = {http://www.pnas.org/content/94/3/814.short; http://www.pnas.org/content/94/3/814.full},
	volume = {94},
	x-fetchedfrom = {Google Scholar},
	year = {1997}
}

@article{elowitz2002stochastic,
  title={Stochastic gene expression in a single cell},
  author={Elowitz, Michael B and Levine, Arnold J and Siggia, Eric D and Swain, Peter S},
  journal={Science},
  volume={297},
  number={5584},
  pages={1183--1186},
  year={2002},
  publisher={American Association for the Advancement of Science}
}

@article{bothma2014dynamic,
  title={Dynamic regulation of eve stripe 2 expression reveals transcriptional bursts in living Drosophila embryos},
  author={Bothma, Jacques P and Garcia, Hernan G and Esposito, Emilia and Schlissel, Gavin and Gregor, Thomas and Levine, Michael},
  journal={Proceedings of the National Academy of Sciences},
  volume={111},
  number={29},
  pages={10598--10603},
  year={2014},
  publisher={National Academy of Sciences}
}

@article{cao2020stochastic,
  title={A stochastic model of gene expression with polymerase recruitment and pause release},
  author={Cao, Zhixing and Filatova, Tatiana and Oyarz{\'u}n, Diego A and Grima, Ramon},
  journal={Biophysical Journal},
  volume={119},
  number={5},
  pages={1002--1014},
  year={2020},
  publisher={Elsevier}
}

@article{shahrezaei2008analytical,
  title={Analytical distributions for stochastic gene expression},
  author={Shahrezaei, Vahid and Swain, Peter S},
  journal={Proceedings of the National Academy of Sciences},
  volume={105},
  number={45},
  pages={17256--17261},
  year={2008},
  publisher={National Academy of Sciences}
}

@article{visco2008exact,
  title={Exact solution of a model DNA-inversion genetic switch with orientational control},
  author={Visco, Paolo and Allen, Rosalind J and Evans, Martin R},
  journal={Physical review letters},
  volume={101},
  number={11},
  pages={118104},
  year={2008},
  publisher={APS}
}

@article{Ham2020,
  author  = {Ham, Lucy and Schnoerr, David and Brackston, Rowan D. and
             Stumpf, Michael P. H.},
  title   = {Exactly solvable models of stochastic gene expression},
  journal = {The Journal of Chemical Physics},
  volume  = {152},
  number  = {14},
  pages   = {144106},
  year    = {2020},
  publisher = {AIP Publishing}
}

@article{kepler2001stochasticity,
  title={Stochasticity in transcriptional regulation: origins, consequences, and mathematical representations},
  author={Kepler, Thomas B and Elston, Timothy C},
  journal={Biophysical journal},
  volume={81},
  number={6},
  pages={3116--3136},
  year={2001},
  publisher={Elsevier}
}

@article{Holehouse2026,
  author  = {Holehouse, James},
  title   = {Quantifying broken detailed balance in transcription},
  journal = {npj Complexity},
  volume  = {3},
  pages   = {10},
  year    = {2026},
  publisher = {Nature Publishing Group},
  doi     = {10.1038/s44260-025-00064-w}
}

@article{zhang2022rate,
  title={The rate of thermodynamic cost against adiabatic and nonadiabatic fluctuations of a single gene circuit in Drosophila embryos},
  author={Zhang, Kun and Ramos, Alexandre Ferreira and Wang, Erkang and Wang, Jin},
  journal={The Journal of Chemical Physics},
  volume={156},
  number={22},
  year={2022},
  publisher={AIP Publishing}
}

@article{cai2006stochastic,
  title={Stochastic protein expression in individual cells at the single molecule level},
  author={Cai, Long and Friedman, Nir and Xie, X Sunney},
  journal={Nature},
  volume={440},
  number={7082},
  pages={358--362},
  year={2006},
  publisher={Nature Publishing Group UK London}
}

@article{kumar2019stochastic,
  title={Stochastic modeling of phenotypic switching and chemoresistance in cancer cell populations},
  author={Kumar, Niraj and Cramer, Gwendolyn M and Dahaj, Seyed Alireza Zamani and Sundaram, Bala and Celli, Jonathan P and Kulkarni, Rahul V},
  journal={Scientific reports},
  volume={9},
  number={1},
  pages={10845},
  year={2019},
  publisher={Nature Publishing Group UK London}
}

@article{sun2020stochastic,
  title={Stochastic gene expression influences the selection of antibiotic resistance mutations},
  author={Sun, Lei and Ashcroft, Peter and Ackermann, Martin and Bonhoeffer, Sebastian},
  journal={Molecular biology and evolution},
  volume={37},
  number={1},
  pages={58--70},
  year={2020},
  publisher={Oxford University Press}
}

@article{Gillespie1977,
  author  = {Gillespie, Daniel T.},
  title   = {Exact stochastic simulation of coupled chemical reactions},
  journal = {The Journal of Physical Chemistry},
  volume  = {81}, number = {25}, pages = {2340--2361}, year = {1977},
  publisher = {American Chemical Society}
}

@book{VanKampen1992,
  author    = {van Kampen, N. G.},
  title     = {Stochastic Processes in Physics and Chemistry},
  edition   = {Revised},
  publisher = {Elsevier},
  year      = {1992},
  address   = {Amsterdam}
}

@article{banerjee2025fano,
  title={Fano Factor as the Key Measure of Sensitivity in Biological Networks},
  author={Banerjee, Kinshuk and Patra, Pintu and Kolomeisky, Anatoly B and Das, Biswajit},
  journal={The Journal of Physical Chemistry Letters},
  volume={16},
  number={39},
  pages={10174--10179},
  year={2025},
  publisher={ACS Publications}
}

@article{Schnakenberg1976,
  author  = {Schnakenberg, J.},
  title   = {Network theory of microscopic and macroscopic behavior of
             master equation systems},
  journal = {Reviews of Modern Physics},
  volume  = {48},
  number  = {4},
  pages   = {571--585},
  year    = {1976},
  publisher = {American Physical Society}
}

@article{mason2011bimodal,
  title={Bimodal distribution of RNA expression levels in human skeletal muscle tissue},
  author={Mason, Clinton C and Hanson, Robert L and Ossowski, Vicky and Bian, Li and Baier, Leslie J and Krakoff, Jonathan and Bogardus, Clifton},
  journal={BMC genomics},
  volume={12},
  number={1},
  pages={98},
  year={2011},
  publisher={Springer}
}

@article{palme2021variation,
  title={Variation in the modality of a yeast signaling pathway is mediated by a single regulator},
  author={Palme, Julius and Wang, Jue and Springer, Michael},
  journal={Elife},
  volume={10},
  pages={e69974},
  year={2021},
  publisher={eLife Sciences Publications, Ltd}
}

@article{mcdavid2014modeling,
  title={Modeling bi-modality improves characterization of cell cycle on gene expression in single cells},
  author={McDavid, Andrew and Dennis, Lucas and Danaher, Patrick and Finak, Greg and Krouse, Michael and Wang, Alice and Webster, Philippa and Beechem, Joseph and Gottardo, Raphael},
  journal={PLoS computational biology},
  volume={10},
  number={7},
  pages={e1003696},
  year={2014},
  publisher={Public Library of Science San Francisco, USA}
}

@article{garcia2016phenotypic,
  title={Phenotypic diversity using bimodal and unimodal expression of stress response proteins},
  author={Garcia-Bernardo, Javier and Dunlop, Mary J},
  journal={Biophysical journal},
  volume={110},
  number={10},
  pages={2278--2287},
  year={2016},
  publisher={Elsevier}
}

@article{baptista2025bimodality,
  title={Bimodality in E. coli gene expression: Sources and robustness to genome-wide stresses},
  author={Baptista, Ines SC and Dash, Suchintak and Arsh, Amir M and Kandavalli, Vinodh and Scandolo, Carlo Maria and Sanders, Barry C and Ribeiro, Andre S},
  journal={PLOS Computational Biology},
  volume={21},
  number={2},
  pages={e1012817},
  year={2025},
  publisher={Public Library of Science San Francisco, CA USA}
}

@article{spratt2026temporal,
  title={Temporal tuning of switch-like virulence expression resolves environmental uncertainty through phenotypic heterogeneity},
  author={Spratt, Madison and Lane, Keara},
  journal={Current Biology},
  volume={36},
  number={13},
  pages={3336--3353},
  year={2026},
  publisher={Elsevier}
}

@article{bu2013microrna,
  title={A microRNA miR-34a-regulated bimodal switch targets Notch in colon cancer stem cells},
  author={Bu, Pengcheng and Chen, Kai-Yuan and Chen, Joyce Huan and Wang, Lihua and Walters, Jewell and Shin, Yong Jun and Goerger, Julian P and Sun, Jian and Witherspoon, Mavee and Rakhilin, Nikolai and others},
  journal={Cell stem cell},
  volume={12},
  number={5},
  pages={602--615},
  year={2013},
  publisher={Elsevier}
}

@article{becskei2001positive,
  title={Positive feedback in eukaryotic gene networks: cell differentiation by graded to binary response conversion},
  author={Becskei, Attila and S{\'e}raphin, Bertrand and Serrano, Luis},
  journal={The EMBO journal},
  volume={20},
  number={10},
  pages={2528--2535},
  year={2001},
  publisher={Springer}
}

@article{friedman2006linking,
  title={Linking stochastic dynamics to population distribution: an analytical framework of gene expression},
  author={Friedman, Nir and Cai, Long and Xie, X Sunney},
  journal={Physical review letters},
  volume={97},
  number={16},
  pages={168302},
  year={2006},
  publisher={APS}
}

@article{portle2007cell,
  title={Cell population heterogeneity in expression of a gene-switching network with fluorescent markers of different half-lives},
  author={Portle, Stephanie and Causey, Thomas B and Wolf, Kim and Bennett, George N and San, Ka-Yiu and Mantzaris, Nikos},
  journal={Journal of biotechnology},
  volume={128},
  number={2},
  pages={362--375},
  year={2007},
  publisher={Elsevier}
}

@article{de2020shape,
  title={The shape of gene expression distributions matter: how incorporating distribution shape improves the interpretation of cancer transcriptomic data},
  author={De Torrent{\'e}, Laurence and Zimmerman, Samuel and Suzuki, Masako and Christopeit, Maximilian and Greally, John M and Mar, Jessica C},
  journal={BMC bioinformatics},
  volume={21},
  number={Suppl 21},
  pages={562},
  year={2020},
  publisher={Springer}
}

@article{hu2018eicirna,
  title={EIciRNA-mediated gene expression: tunability and bimodality},
  author={Hu, Qi and Zhou, Tianshou},
  journal={FEBS letters},
  volume={592},
  number={20},
  pages={3460--3471},
  year={2018},
  publisher={Wiley Online Library}
}

@article{roy2021persister,
  title={Persister control by leveraging dormancy associated reduction of antibiotic efflux},
  author={Roy, Sweta and Bahar, Ali Adem and Gu, Huan and Nangia, Shikha and Sauer, Karin and Ren, Dacheng},
  journal={PLoS Pathogens},
  volume={17},
  number={12},
  pages={e1010144},
  year={2021},
  publisher={Public Library of Science San Francisco, CA USA}
}

@Article{Pu2016,
  author    = {Pu, Yingying and Zhao, Zhilun and Li, Yingxing and Zou, Jin and Ma, Qi and Zhao, Yanna and Ke, Yuehua and Zhu, Yun and Chen, Huiyi and Baker, Matthew AB and others},
  title     = {Enhanced efflux activity facilitates drug tolerance in dormant bacterial cells},
  journal   = {Molecular cell},
  year      = {2016},
  volume    = {62},
  number    = {2},
  pages     = {284--294},
  publisher = {Elsevier},
}

@article{shine2024co,
  title={Co-transcriptional gene regulation in eukaryotes and prokaryotes},
  author={Shine, Morgan and Gordon, Jackson and Sch{\"a}rfen, Leonard and Zigackova, Dagmar and Herzel, Lydia and Neugebauer, Karla M},
  journal={Nature Reviews Molecular Cell Biology},
  volume={25},
  number={7},
  pages={534--554},
  year={2024},
  publisher={Nature Publishing Group UK London}
}

@article{trinquier2023effect,
  title={Effect of tRNA maturase depletion on levels and stabilities of ribosome assembly cofactor and other mRNAs in Bacillus subtilis},
  author={Trinquier, Aude and Condon, Ciar{\'a}n and Braun, Fr{\'e}d{\'e}rique},
  journal={Microbiology Spectrum},
  volume={11},
  number={2},
  pages={e05134--22},
  year={2023},
  publisher={American Society for Microbiology 1752 N St., NW, Washington, DC}
}

@article{kim2026spatial,
  title={Spatial and genetic constraints govern transcription--translation coupling and mRNA degradation in bacteria},
  author={Kim, Seunghyeon and Zhang, Yan and Ju, Xiangwu and Hassan, Albur and Wang, Yu-Huan and Liu, Shixin and Kim, Sangjin},
  journal={Nature Microbiology},
  pages={1--14},
  year={2026},
  publisher={Nature Publishing Group UK London}
}

@article{richards2021riboswitch,
  title={Riboswitch control of bacterial RNA stability},
  author={Richards, Jamie and Belasco, Joel G},
  journal={Molecular microbiology},
  volume={116},
  number={2},
  pages={361--365},
  year={2021},
  publisher={Wiley Online Library}
}

@article{vargas2020regulation,
  title={Regulation of mRNA stability during bacterial stress responses},
  author={Vargas-Blanco, Diego A and Shell, Scarlet S},
  journal={Frontiers in microbiology},
  volume={11},
  pages={2111},
  year={2020},
  publisher={Frontiers Media SA}
}

@article{dattani2017stochastic,
  title={Stochastic models of gene transcription with upstream drives: exact solution and sample path characterization},
  author={Dattani, Justine and Barahona, Mauricio},
  journal={Journal of The Royal Society Interface},
  volume={14},
  number={126},
  pages={20160833},
  year={2017}
}

@article{park2018chemical,
  title={The chemical fluctuation theorem governing gene expression},
  author={Park, Seong Jun and Song, Sanggeun and Yang, Gil-Suk and Kim, Philip M and Yoon, Sangwoon and Kim, Ji-Hyun and Sung, Jaeyoung},
  journal={Nature communications},
  volume={9},
  number={1},
  pages={297},
  year={2018},
  publisher={Nature Publishing Group UK London}
}

\end{document}